\documentclass[aps,prx,twocolumn]{revtex4}
\usepackage{amsfonts}
\usepackage{amsmath}
\usepackage{graphicx}
\usepackage{dsfont}
\usepackage{diagbox}
\usepackage{array}
\usepackage{amssymb}
\usepackage{bbm}
\usepackage{float}
\usepackage{subfigure}
\usepackage{url}
\usepackage{hyperref}
\usepackage{bm}

\newcommand{\PreserveBackslash}[1]{\let\temp=\\#1\let\\=\temp}
\newcolumntype{C}[1]{>{\PreserveBackslash\centering}p{#1}}
\newcolumntype{R}[1]{>{\PreserveBackslash\raggedleft}p{#1}}
\newcolumntype{L}[1]{>{\PreserveBackslash\raggedright}p{#1}}

\begin{document}

\title{Constructing tensor network wavefunction for a generic
two-dimensional quantum phase transition via thermofield double states}
\author{Wen-Tao Xu$^{1}$ and Guang-Ming Zhang$^{1,2}$}
\affiliation{$^{1}$State Key Laboratory of Low-Dimensional Quantum Physics and Department
of Physics, Tsinghua University, Beijing 100084, China. \\
$^{2}$Frontier Science Center for Quantum Information, Beijing 100084, China.}
\date{\today}

\begin{abstract}
The most important feature of two-dimensional quantum Rokhsar-Kivelson (RK)
type models is that their ground state wavefunction norms can be mapped into
the partition functions of two-dimensional statistical models so that the
quantum phase transitions become the thermal phase transitions of the
corresponding statistical models. For a generic quantum critical point, we
generalize the framework of RK wavefunctions by introducing the concept of
the thermofield double (TFD) state, which is a purification of the
equilibrium density operator. Moreover, by expressing the TFD state in terms
of the projected entangled pair state, its $N$-order of R\'{e}nyi entropy
results in a three-dimensional statistical model in Euclidian spacetime,
describing the generic quantum phase transitions. Using the toric code model
with two parallel magnetic fields as an example, we explain these ideas and
derive the partition function of the three-dimensional $Z_2$ lattice
gauge-Higgs model, where the phase transitions are characterized by the
three-dimensional universality classes.
\end{abstract}

\maketitle

\textit{Introduction}. -Recently quantum spin liquids, topological phases
of matter, and their quantum phase transitions have been attracted enormous
research interest, and there has been a very active development of tensor
network states to study these correlated quantum many-body phenomena\cite%
{Verstraete-Adv-Phys,Schollwock-Ann-Phys,Orus-Ann-Phys,Haegeman-Verstraete2017}%
. Since the ground states and low-energy excited states of two-dimensional
(2D) quantum systems with local interactions are only weak entangled, the
entanglement entropy of a region scales with its area\cite%
{Wolf-Cirac-2008,Eisert-rmp-2010}. Then infinite projected entangled-pair
states (iPEPS) or tensor network states\cite%
{Verstraete-Cirac-2004,Jordan-prl-2008,Nishino-2001,Nishino-2004} has become
a powerful theoretical tool to characterize those correlated quantum
many-body phases in the thermodynamic limit\cite%
{Schuch-prb-2012,Schuch-prl-2013}. However, there still exists an open
problem how well the iPEPS can describe two-dimensional quantum critical
states emerging at the quantum critical points\cite%
{Lauchli-prx-2018,Corboz-prx-2018}.

Since the norm of the iPEPS can be related to a partition function of
classical statistical models, the zero-temperature quantum phase transitions
can be mapped to the temperature-driven classical phase transitions\cite%
{Isakov-2011,GaugingTNS_2015,zhu_gapless_2019,Xu_fib_2020}, and there is a
class of exact 2D critical iPEPS with a small value of bond dimension,
describing the ground states of the generalized Rokhsar-Kivelson (RK)
Hamiltonians at their critical point\cite%
{ardonne_2004,castelnovo-2005,Verstraete_2006}. However, a generic 2D
quantum phase transition should be described by the partition function of a
3D classical statistical model in the Euclidian space, where an extra
dimension corresponds to the "imaginary time" to represent thermal
fluctuations at the critical point. And the emerging critical properties of
the 2D quantum model belong to the same universality class as that of the 3D
classical system. Therefore, the most important question is to establish a
general framework connecting the tensor network wavefunction for a 2D
quantum system to a 3D classical partition function, and to find out the
necessary conditions under which the general framework can be applied.

In this paper, we generalize the RK wavefunctions to the framework of the
so-call thermofield double (TFD) states\cite{THERMO_1996}, which are defined
in an enlarged Hilbert space $\mathcal{H}_{P}\otimes \mathcal{H}_{F}$
consisting of the physical space $\mathcal{H}_{P}$ and a fictitious space $%
\mathcal{H}_{F}$. The TFD state is a purification of a mixed density
operator in thermal equilibrium, and the thermal fluctuations is encoded in
the form of quantum entanglement between the physical and fictitious parts.
For a class of 2D quantum systems, their model Hamiltonians can be
divided into two noncommutative parts $\hat{H}=\hat{H}^{(1)}+\hat{H}^{(2)}$
and each part is a sum of the commuting local terms. We can introduce a
tensor network TFD state with variational parameters, and then a reduced TFD
density operator can be obtained by integrating out the fictitious degrees
of freedom. We prove that the $N$-th order R\'{e}nyi entropy gives rise to
the partition function of 3D statistical models in the Euclidian space,
where the imaginary time dimension is discretized into $N$ layers. In the
large $N$ limit, we expect that the phase transitions of the 3D statistical
models characterize the quantum phase transitions in the 2D quantum model.
As an example, we carefully study the toric code model in two parallel fields%
\cite{kitaev_toric_code_2003,Vidal2009,Gauge_Higgs_2010}. By constructing a
variational tensor network TFD state, we successfully derive the partition
function of the 3D lattice gauge-Higgs model\cite%
{Fradkin_1979,Gauge_Higgs_1980,Gauge_Higgs_2010}, which can be used to
describe the topological quantum phase transitions out of the toric code
phase.

\textit{General framework}. -From a 2D classical Hamiltonian $E(\sigma
_{1},\cdots ,\sigma _{n})$ with spin variables $\{\sigma _{i}\}$, the
RK-type wavefunction with a parameter $\beta $ is defined by%
\begin{equation}
|\psi (\beta )\rangle =\sum_{\bm{\sigma}}\text{e}^{-\beta E(\bm{\sigma})/2}|%
\bm{\sigma}\rangle .  \label{RK}
\end{equation}%
where $|\bm{\sigma}\rangle =|\sigma _{1},\cdots ,\sigma _{n}\rangle $ and $%
\{|\bm{\sigma}\rangle \}$ is a set of completely orthogonal basis. The
wavefunction norm precisely corresponds to the partition function of a
classical model with $\beta $ as the inverse temperature:
\begin{equation}
Z(\beta )=\langle \psi (\beta )|\psi (\beta )\rangle =\sum_{\bm{\sigma}}\exp
[-\beta E(\bm{\sigma})].
\end{equation}%
With such a quantum-classical mapping, all equal-time correlation functions of 
local observable operators are mapped into the correlation functions of the 
classical partition function. Then the quantum phase transitions in the RK 
wavefunction are described by the partition function of a 2D classical model.

When we consider a 2D quantum Hamiltonian $\hat{H}(\hat{\sigma}_{i},\cdots ,%
\hat{\sigma}_{n})$ with spin operators $\{\hat{\sigma}_{i}\}$, a generic quantum 
wavefunction function can be expressed as
\begin{equation}
|\Psi _{0}(\beta )\rangle =\sum_{\bm{\sigma}}e^{-\beta \hat{H}(\hat{\sigma}%
_{i}\cdots \hat{\sigma}_{n})/2}|\bm{\sigma}\rangle .  \label{RK_prime}
\end{equation}%
When $\beta \rightarrow \infty $, the wavefunction $|\Psi _{0}(\infty
)\rangle $ approaches to the ground state of $\hat{H}$. Unlike the RK type
wavefunctions, the norm of the wavefunction $\langle \Psi _{0}(\beta )|\Psi
_{0}(\beta )\rangle $ is significantly different from the partition function
of the quantum Hamiltonian $\mathcal{Z}(\beta )=\text{tr}e^{-\beta \hat{H}}$%
. The former sums over both diagonal and off-diagonal matrix elements of the
density operator $e^{-\beta \hat{H}}$, the later is just a sum of diagonal
matrix elements. So we can not make the norm of a pure state equivalent to
the trace of a mixed state in the original Hilbert space, except the gapped
system at zero temperature. However, in an enlarged Hilbert space\cite%
{THERMO_1996}, it can be done by constructing the so-called TFD state:
\begin{equation}
|\tilde{\Psi}_{0}(\beta )\rangle =\sum_{\bm{\sigma}}\left[ e^{-\beta \hat{H}%
/2}|\bm{\sigma}\rangle _{P}\right] \otimes |\bm{\sigma}\rangle _{F},
\label{TFD}
\end{equation}%
where the enlarged Hilbert space $\mathcal{H}_{P}\otimes \mathcal{H}_{F}$
consists of the physical space $\mathcal{H}_{P}$ and a fictitious one $%
\mathcal{H}_{F}$, and the square root of the density operator acts on the
physical space only. It is always sufficient to choose $\mathcal{H}_{F}$ to
be identical to $\mathcal{H}_{P}$, \textquotedblleft
doubling\textquotedblright\ each degree of freedom. For a gapped quantum
system with a non-degenerate ground state, the entanglement between the
physical and fictitious parts decreases with increasing $\beta $. In the
limit of $\beta \rightarrow \infty $, the density operator becomes a ground
state projector, leading to a product state $|\tilde{\Psi}%
_{0}(\infty)\rangle =|\Psi _{0}(\infty )\rangle _{P}\otimes |\Psi _{0}^{\ast
}(\infty )\rangle _{F}$. So for a gapped system at zero temperature, using
the TFD state is not necessary. However, it is essential for a critical
system to consider the entanglement between two parts of a TFD state.

Then tracing out the degrees of freedom in the fictitious space yields a
reduced density matrix of the TFD state
\begin{equation}
\rho _{0}=\text{tr}_{F}|\tilde{\Psi}_{0}(\beta )\rangle \langle \tilde{\Psi}%
_{0}(\beta )|=e^{-\beta \hat{H}},
\end{equation}%
which is the exact Gibbs density operator. So the information of the thermal
fluctuations is just encoded in the form of the quantum entanglement between
the physical and fictitious parts, and the norm of the TFD state corresponds 
to the partition function of the 2D quantum Hamiltonian:
\begin{equation}
\mathcal{Z}(\beta )=\langle \tilde{\Psi}_{0}(\beta )|\tilde{\Psi}_{0}(\beta
)\rangle =\text{tr}e^{-\beta \hat{H}}.
\end{equation}%
Therefore, the TFD state reconciles the norm and trace operations and
generalizes the RK-type wavefunctions, and the unequal time correlation
function can be calculated with the TFD states.

Usually it is very hard to express the wavefunction coefficients as the
products of local Boltzmann weights\cite%
{Verstraete_2006,Approximating_Gibbs_2015}. For a special class of quantum
systems, however, their model Hamiltonians can be divided into two
noncommutative parts $\hat{H}=\hat{H}^{(1)}+\hat{H}^{(2)}$ and each part is
a sum of the commuting local terms. Then a good variational ground state
wavefunction can be expressed as\cite{Bridging2017}
\begin{equation}
|\Psi (\alpha _{1},\alpha _{2})\rangle =\sum_{\bm{\sigma}}e^{-\alpha _{1}%
\hat{H}^{(1)}/2}e^{-\alpha _{2}\hat{H}^{(2)}/2}|\bm{\sigma}\rangle ,
\label{variational_TFD}
\end{equation}%
where $(\alpha_{1},\alpha_{2})$ are two variational parameters. Then we can
express the corresponding TFD state as
\begin{equation}
|\tilde{\Psi}(\alpha _{1},\alpha _{2})\rangle =\sum_{\bm{\bar{\sigma}},\bm{\sigma}}
(e^{-\alpha_1\hat{H}^{(1)}/2}e^{-\alpha_2\hat{H}^{(2)}/2})_{\bar{\sigma}{\sigma}}
|\bm{\bar{\sigma}}\rangle _{P}\otimes |\bm{\sigma}\rangle _{F},
\label{Main}
\end{equation}%
and those coefficients can be easily written as the products of local
Boltzmann weights
\begin{equation*}
(e^{-\alpha_1\hat{H}^{(1)}/2}e^{-\alpha_2\hat{H}^{(2)}/2})_{\bar{\sigma}{\sigma}}
=\sum_{\bm{s}}\prod_{v}
\exp \left[ {-\epsilon _{s_{v}}^{\sigma_{v},\bar{\sigma}_{v}}(\alpha _{1},\alpha _{2})}\right] ,
\end{equation*}%
where the auxiliary degrees of freedom $\bm{s}=s_{1},\cdots ,s_{m}$ have
been introduced and $\sigma _{v},s_{v},\bar{\sigma}_{v}$ are the sets of
spin variables belonging to the vertex $v$. When the local Boltzmann weights
are replaced by local tensors, the TFD state wavefunction becomes an iPEPS
with physical degrees of freedom $\bm{\bar{\sigma}},\bm{\sigma}$.

From Eq. (\ref{Main}), after integrating out the fictitious degrees of
freedom, we obtain a reduced TFD density operator
\begin{equation}
\rho =\text{tr}_{F}|\tilde{\Psi}\rangle \langle \tilde{\Psi}|=e^{-\alpha _{1}%
\hat{H}^{(1)}/2}e^{-\alpha _{2}\hat{H}^{((2)}}e^{-\alpha _{1}\hat{H}%
^{(1)}/2}.
\end{equation}%
In order to construct the partition function of a 3D statistical model for
the critical iPEPS, we stack $N$ copies of $\rho $ and contract all degrees
of freedom including those at the top of the $N$-th layer and those at the
first layer. Then we identify $\mathcal{Z}_{N}=$tr$\rho ^{N}$ as a partition
function in the Euclidian spacetime. Actually the usual $N$-th order R\'{e}%
nyi entropy is given by
\begin{equation}
S_{N}=\frac{\log \text{tr}\rho ^{N}}{1-N}.
\end{equation}%
So the derived partition function $\mathcal{Z}_{N}$ can be explicitly
written in terms of the local tensors of the iPEPS,
\begin{equation}
\mathcal{Z}_{N}=\sum_{\substack{ \bm{\sigma}_{1}^{[1]}\cdots \bm{\sigma}%
_{n}^{[N]},  \\ \bm{s}_{1}^{[1]}\cdots \bm{s}_{1}^{[N]}}}\exp \left[ {%
-\sum_{\tau =1}^{N}\sum_{v}\epsilon _{s_{v}^{[\tau ]}}^{\sigma _{v}^{[\tau
]},\sigma _{v}^{[\tau +1]}}(2\alpha _{1},2\alpha _{2})}\right] ,
\end{equation}%
where $\tau $ denotes the discretized imaginary time. When $N$ is sufficient
large, the quantum-classical mapping of the RK-type wavefunction is thus
generalized by using the concept of the TFD state via quantum entanglement,
and the partition function of a 2D quantum system is transformed into that
of a 3D classical one, which can be simulated with the classical resources.
If the intra-layer interactions in the spatial directions have the same form
as the inter-layer interactions in the imaginary time direction, the
isotropic 3D model at the critical point has an emergent Lorentz symmetry
and the dynamical critical exponent $z=1$.

\textit{Toric code model in two parallel fields}. -To illustrate the above
general theory, we consider the toric code model defined on a square lattice
\begin{eqnarray}
H_{\text{TC}} &=&-\sum_{v}A_{v}-\sum_{p}B_{p}, \\
A_{v} &=&-\sum_{v}\prod_{\langle ij\rangle \in v}\sigma _{ij}^{x},\quad
B_{p}=-\sum_{p}\prod_{\langle ij\rangle \in p}\sigma _{ij}^{z},  \notag
\end{eqnarray}%
where the vertex and plaquette terms $A_{v}$ and $B_{p}$ involve four Pauli
spin-1/2 operators located on the bonds between sites $i$ and $j$. $A_{v}=-1$
is associated with an electric charge excitation, while $B_{p}=-1$ is
associated with a magnetic flux excitation. The ground states can be
projected out by the projectors of the subspaces:
\begin{equation*}
|\Psi _{\text{TC}}^{m}\rangle =\prod_{v}\left( 1+A_{v}\right) |\uparrow
\rangle ^{\otimes L},\text{ }|\Psi _{\text{TC}}^{e}\rangle =\prod_{p}\left(
1+B_{p}\right) |+\rangle ^{\otimes L}.
\end{equation*}%
where $|+\rangle =(|\uparrow \rangle +|\downarrow \rangle )/\sqrt{2}$ and $L$
is the bond number. On a manifold with the trivial topology, the ground
state is non-degenerate, $|\Psi _{\text{TC}}^{e}\rangle =|\Psi _{\text{TC}%
}^{m}\rangle $.

\begin{figure}[tbp]
\centering
\includegraphics[width=0.45\textwidth]{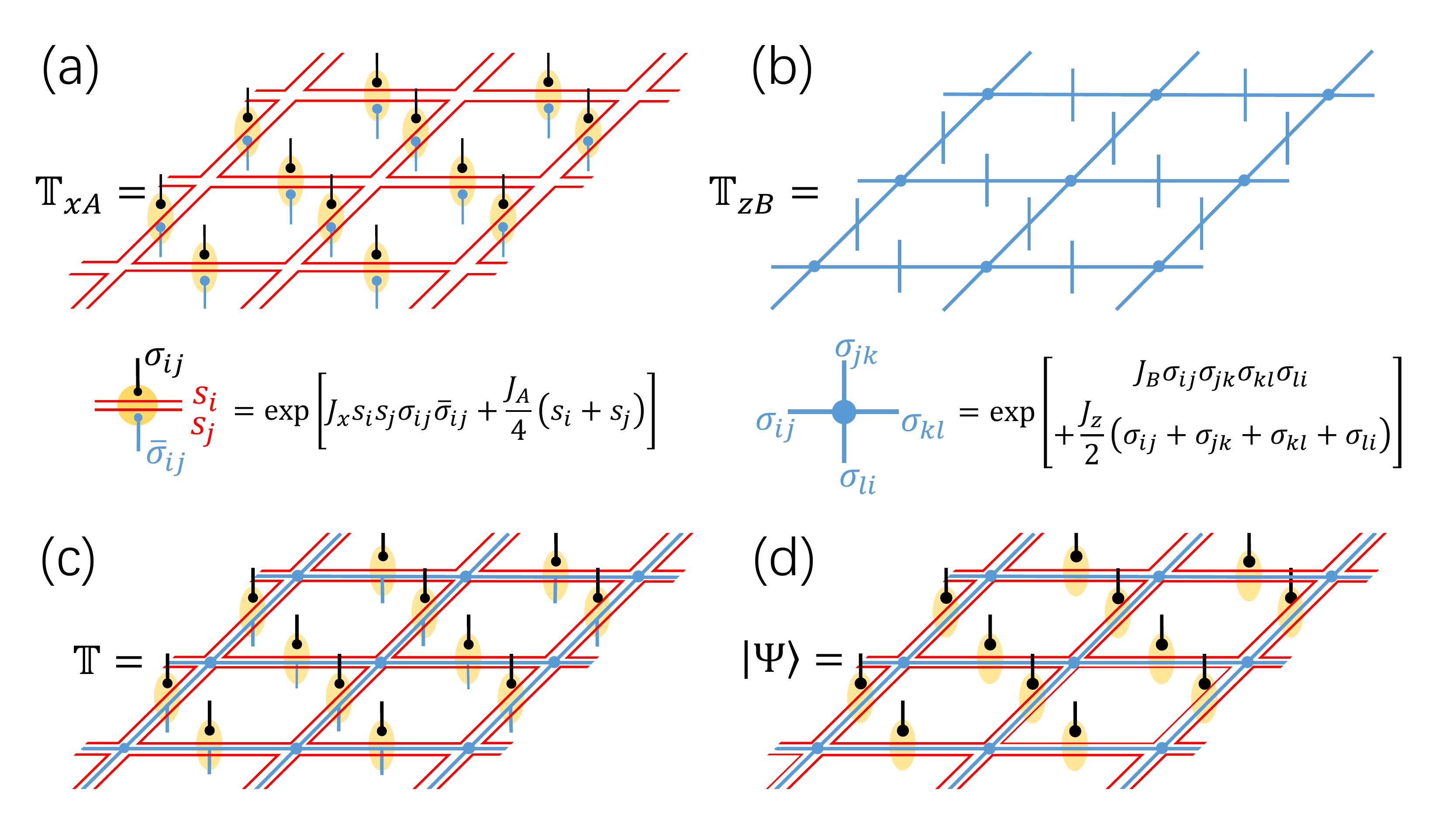}
\caption{(a) The double-line PEPO representation of $\mathbb{T}_{xA}$ and a
local tensor on a bond. When this PEPO is reduced to an off-diagonal
projector, the local tensor becomes $({1+s_{i}s_{j}\protect\sigma _{ij}\bar{%
\protect\sigma}_{ij}})/2$. (b) The single-line PEPO representation of $%
\mathbb{T}_{zB}$ and a local tensor on the vertex $\hat{v}$. When this PEPO
is reduced to the diagonal projector, the local tensor becomes $1+\protect%
\sigma _{ij}\protect\sigma _{jk}\protect\sigma _{kl}\protect\sigma _{li}$.
(c) The triple-line PEPO representation of $\mathbb{T}$. (d) The triple-line
PEPS of $|\Psi (h_{x},a,h_{z},b)\rangle $.}
\label{TN}
\end{figure}

It is useful to represent the ground states as iPEPS. When auxiliary degrees
of freedom $s_{j}=\pm 1$ on each site is introduced, the off-diagonal
projector can be expressed as a projected entangled pair operator (PEPO)
\begin{equation*}
\prod_{v}\left( 1+A_{v}\right) =\sum_{\bm{s}\bm{\sigma}\bm{\bar{\sigma}}%
}\prod_{ij}\frac{1+s_{i}s_{j}\sigma _{ij}\bar{\sigma}_{ij}}{2}|\bm{\sigma}%
\rangle \langle \bm{\bar{\sigma}}|,
\end{equation*}%
corresponding to a double-line PEPO on the \emph{dual} lattice in Fig. \ref%
{TN} (a), while the diagonal projector is just a single line PEPO on the
dual lattice
\begin{equation*}
\prod_{p}\left( 1+B_{p}\right) =\sum_{\bm{\sigma}}\prod_{\hat{v}}\left[
1+\prod_{\langle ij\rangle \in \hat{v}}\sigma _{ij}\right] |\bm{\sigma}%
\rangle \langle \bm{\sigma}|,  \label{diagonal}
\end{equation*}
with $\hat{v}=p$ as a vertex of the dual lattice, as shown in Fig. \ref{TN}(b).

In order to study topological quantum phase transitions out of the
topological phase, two different parallel fields are introduced into the
toric code model
\begin{equation}
H=H_{TC}-\mathfrak{h}_{x}\sum_{\langle ij\rangle }\sigma _{ij}^{x}-\mathfrak{%
h}_{z}\sum_{\langle ij\rangle }\sigma _{ij}^{z}.  \label{HTC_field}
\end{equation}%
This generalized toric code model can not be exactly solved, because the
eigenvalues of $A_{v}$ and $B_{p}$ are no longer good quantum numbers.
According to our general proposal, a good variational wavefunction can be
expressed as
\begin{eqnarray}
&&|\Psi \left( h_{x},a,h_{z},b\right) \rangle =\mathbb{T}_{xA}\mathbb{T}%
_{zB}|+\rangle ^{\otimes L},\notag \\
&&\mathbb{T}_{xA}=\prod_{\langle ij\rangle }\left( 1+h_{x}\sigma
_{ij}^{x}\right) \prod_{v}\left( 1+aA_{v}\right) ,  \notag \\
&&\mathbb{T}_{zB}=\prod_{\langle ij\rangle }\left( 1+h_{z}\sigma
_{ij}^{z}\right) \prod_{p}\left( 1+bB_{p}\right) ,
\end{eqnarray}%
with the parameters $h_{x},h_{z},a,b\in \lbrack 0,1]$. When $a=0$, $b=1$,
this wavefunction is reduced to the deformed wavefunction\cite%
{GaugingTNS_2015}, whose norm can be mapped into the partition function of
the 2D classical Ashkin-Teller model\cite{zhu_gapless_2019}. When $h_{z}=0$,
this variational wavefunction has been previously used to consider the
topological phase transitions of the toric code phase\cite{Schotte2019}.

To represent the general variational wavefunction as the tensor networks,
the off-diagonal and diagonal deformation operators are written as PEPOs
\begin{eqnarray}
\mathbb{T}_{xA} &\propto &\sum_{\bm{s}\bm{\sigma}\bar{\bm{\sigma}}}\exp %
\left[ {J_{A}\sum_{i}s_{i}+J_{x}\sum_{\langle ij\rangle }s_{i}s_{j}\sigma
_{ij}\bar{\sigma}_{ij}}\right] |\bm{\sigma}\rangle \langle \bar{\bm{\sigma}}%
|,  \notag \\
\mathbb{T}_{zB} &\propto &\sum_{\bm{\sigma}}\exp \left[ {J_{z}\sum_{\langle
ij\rangle }\sigma _{ij}+J_{B}\sum_{p}\prod_{\langle ij\rangle \in p}\sigma
_{ij}}\right] |\bm{\sigma}\rangle \langle \bm{\sigma}|.  \notag
\end{eqnarray}%
where $2J_{A}=-\log a$, $J_{z}=\tanh ^{-1}h_{z}$, $2J_{x}=-\log h_{x}$, $%
J_{B}=\tanh ^{-1}b$. Then $\mathbb{T}=\mathbb{T}_{xA}\mathbb{T}_{zB}$ is a
triple-line PEPO as displayed in Fig. \ref{TN} (c)%
\begin{eqnarray}
\mathbb{T} &\propto &\sum_{\bm{\sigma}\bm{\bar\sigma}\bm{s}}\exp \left(
J_{A}\sum_{i}s_{i}+J_{x}\sum_{\langle ij\rangle }s_{i}s_{j}\sigma _{ij}\bar{%
\sigma}_{ij}\right.  \notag \\
&&\text{ \ \ }\left. +J_{z}\sum_{\langle ij\rangle }\sigma
_{ij}+J_{B}\sum_{p}\prod_{\langle ij\rangle \in p}\bar{\sigma}_{ij}|\right) %
\bm{\sigma}\rangle \langle \bm{\bar{\sigma}}|.
\end{eqnarray}%
Acting $\mathbb{T}$ on the reference state $|+\rangle ^{\otimes L}$ gives
rise to the triple-line PEPS $|\Psi \left( h_{x},a,h_{z},b\right) \rangle $,
displayed in Fig.\ref{TN}(d).

\begin{figure}[tbp]
\centering
\includegraphics[width=0.35\textwidth]{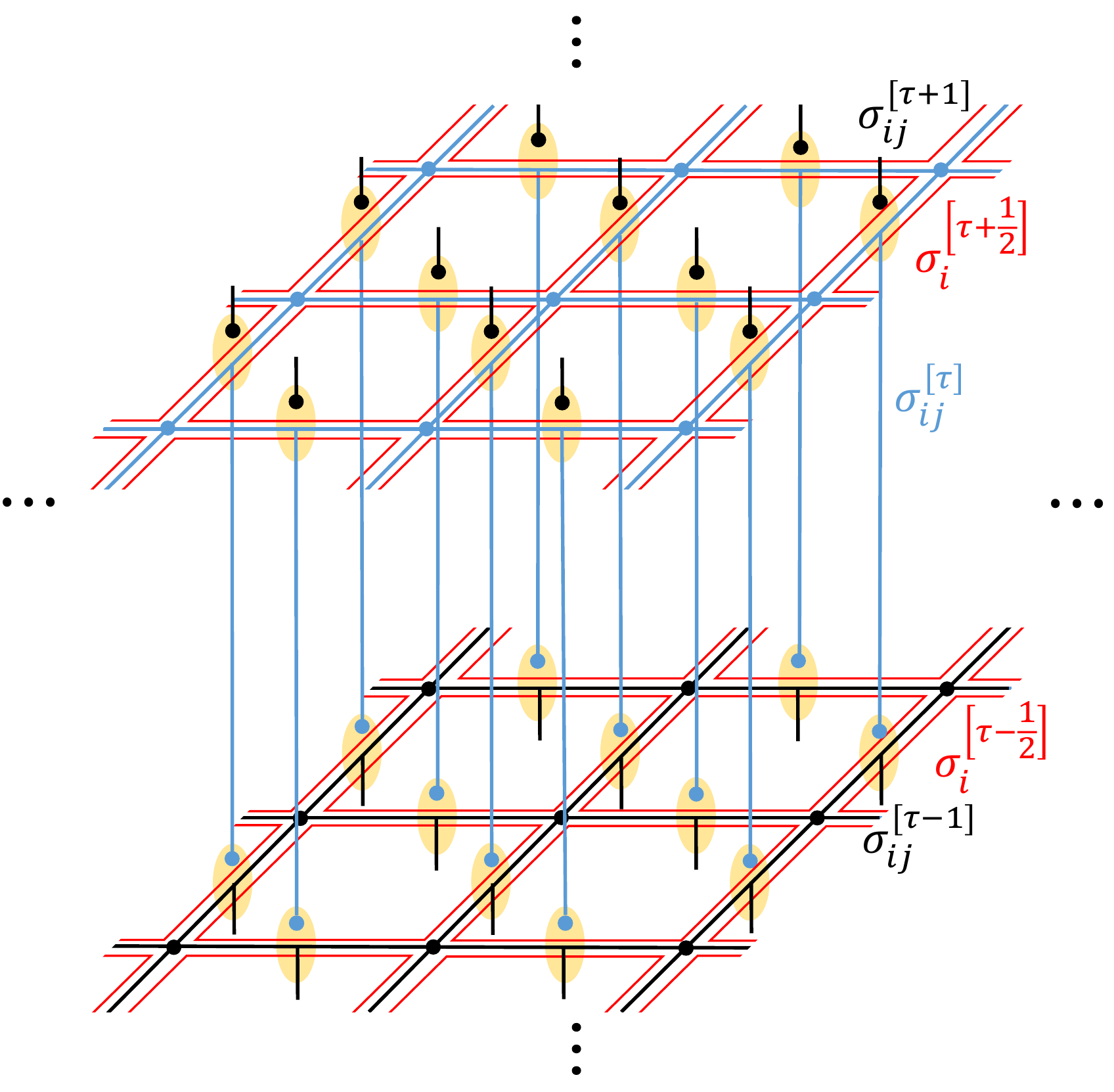}
\caption{The entanglement between two neighbouring layers in a
three-dimensional tensor network structure. }
\label{RDM}
\end{figure}

With the operator $\mathbb{T}$, the corresponding TFD state can be constructed as
\begin{equation}
|\tilde{\Psi}\left( h_{x},a,h_{z},b\right) \rangle =\sum_{\bm{\sigma}}%
\mathbb{T}|\bm{\sigma}\rangle _{P}\otimes |\bm{\sigma}\rangle _{F}.
\end{equation}%
Then a reduced density matrix can be obtained by tracing out the degrees of
freedom of the fictitious space:
\begin{equation*}
\rho =\text{tr}_{F}|\tilde{\Psi}\left( h_{x},a,h_{z},b\right) \rangle
\langle \tilde{\Psi}\left( h_{x},a,h_{z},b\right) |=\mathbb{T}\mathbb{T}%
^{\dagger }.
\end{equation*}%
Actually $\mathbb{T}_{xA}$ can be regarded as a similar transformation, so
we can further simplified the reduced density operator as
\begin{equation}
\rho \sim \mathbb{T}(\bar{J}_{A},\bar{J}_{x},\bar{J}_{z},\bar{J}_{B}).
\end{equation}%
with the modified parameters
\begin{eqnarray*}
\bar{J}_{A} &=&\frac{1}{2}\log \frac{1+a^{2}}{2a},\quad \quad \bar{J}%
_{z}=2\tanh ^{-1}h_{z},\quad  \\
\bar{J}_{x} &=&\frac{1}{2}\log \frac{1+h_{x}^{2}}{2h_{x}},\quad \quad \bar{J}%
_{B}=2\tanh ^{-1}b.
\end{eqnarray*}%
According to the general framework, the partition function of a 3D classical
model can be constructed from the R\'{e}nyi entropy:
\begin{equation}
\mathcal{Z}_{N}=\text{tr}\rho ^{N}\propto \text{tr}\mathbb{T}^{N}(\bar{J}%
_{A},\bar{J}_{x},\bar{J}_{z},\bar{J}_{B}).
\end{equation}%
By ignoring the unimportant coefficients, the explicit form of the partition
function can be written as
\begin{eqnarray*}
&&\mathcal{Z}_{N} =\sum_{\substack{ \bm{\sigma}^{[1]},\bm{\sigma}^{[\frac{1}{%
2}]},\cdots ,  \\ \bm{\sigma}^{[N+\frac{1}{2}]},\bm{\sigma}^{[N]}}}%
\prod_{\tau =1}^{N}\exp \left[ \bar{J}_{A}\sum_{i}\sigma _{i}^{[\tau +\frac{1%
}{2}]}+\bar{J}_{z}\sum_{\langle ij\rangle }\sigma _{ij}^{[\tau ]}\right. \\
&&\left. +\bar{J}_{x}\sum_{\langle ij\rangle }\sigma _{i}^{[\tau +\frac{1}{2}%
]}\sigma _{j}^{[\tau +\frac{1}{2}]}\sigma _{ij}^{[\tau ]}\sigma _{ij}^{[\tau
+1]}+\bar{J}_{B}\sum_{p}\prod_{\langle ij\rangle \in p}\sigma _{ij}^{[\tau ]}%
\right] ,
\end{eqnarray*}%
where we have substituted the labels $\sigma _{ij}^{[\tau ]},\sigma
_{ij}^{[\tau +1]},\sigma _{i}^{[\tau +\frac{1}{2}]}$ for the labels $\sigma
_{ij},\bar{\sigma}_{ij},s_{i}$. The partition function $\mathcal{Z}_{N}$ is
defined in the 3D Euclidian spacetime in the large $N$ limit, the entanglement 
structure between different layers is displayed in the Fig.\ref{RDM}. 

\begin{figure}[tbp]
\centering
\includegraphics[width=0.45\textwidth]{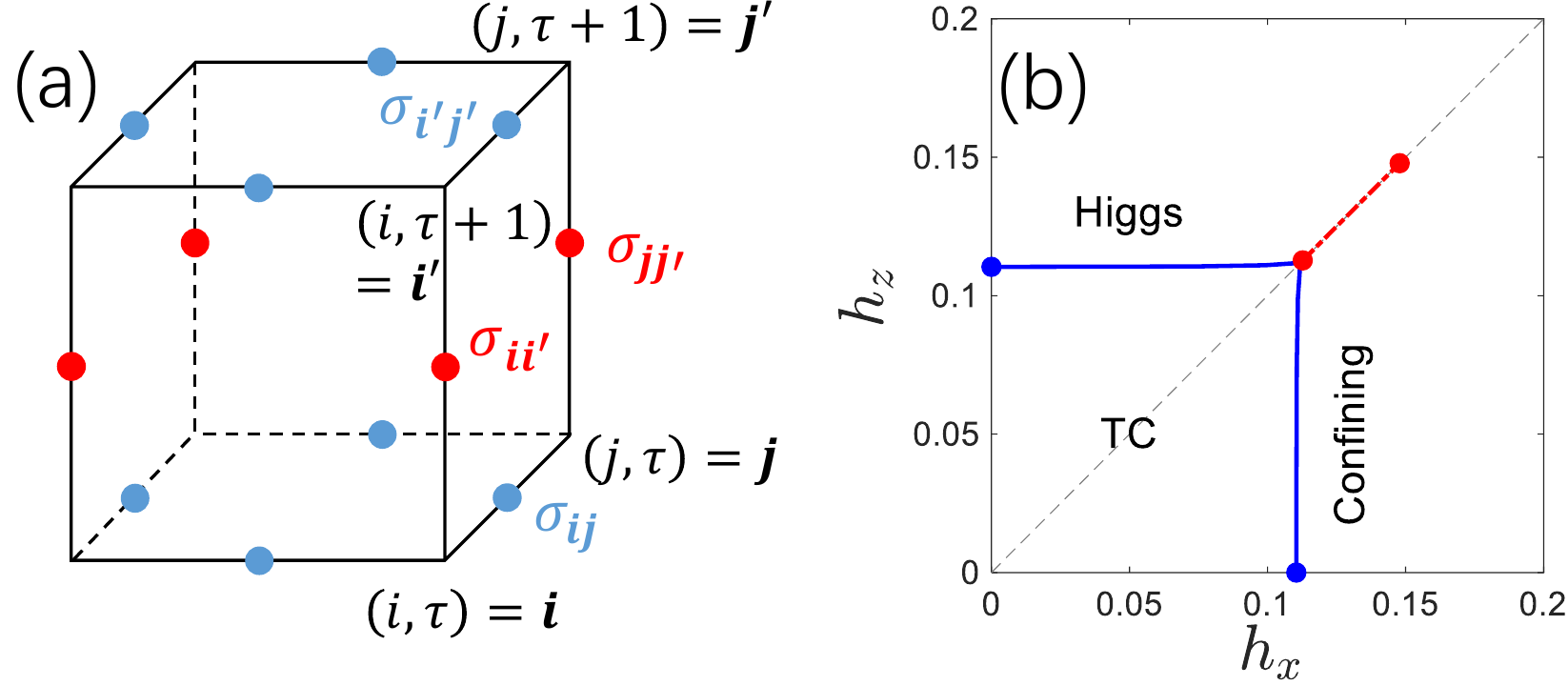}
\caption{(a) The unit cell of the three-dimensional cubic lattice. The
degrees of freedom of Ising gauge Higgs model locate on the bonds of the
lattice. Each unit contains three plaquettes in average, each plaquette
corresponds to a four-spin interaction term. (b) The phase diagram of the 3D
isotropic Ising gauge Higgs model.}
\label{Gauge_higgs}
\end{figure}

When the isotropic inter-layer and intra-layer interactions are required, we
have to restrict the variational parameters as
\begin{eqnarray*}
\bar{J}_{A} &=&\bar{J}_{z}=2\tanh ^{-1}h_{z}\equiv K,  \notag \\
\bar{J}_{x} &=&\bar{J}_{B}=\frac{1}{2}\log \frac{1+h_{x}^{2}}{2h_{x}}\equiv
K_{\bm{p}},
\end{eqnarray*}%
and the number of the varying parameters are reduced. Then the obtained
partition function can be simplified as
\begin{equation}
\mathcal{Z}_{N}=\sum_{\bm{\sigma}}\exp \left[ K\sum_{\langle \bm{i}\bm{j}%
\rangle }\sigma _{\bm{i}\bm{j}}+K_{\bm{p}}\sum_{\bm{p}}\prod_{\langle \bm{i}%
\bm{j}\rangle \in \bm{p}}\sigma _{\bm{i}\bm{j}}\right] ,
\end{equation}%
where $\bm{i}=(i,\tau )$ denotes the cubic lattice sites, $\bm{p}$ are the
plaquettes of the cubic lattice, and $\bm{\sigma}$ stands for the
configurations of all spins $\sigma _{\bm{ij}}$ on the bonds of the cubic
lattice, as displayed in Fig. \ref{Gauge_higgs} (a). Actually this partition
function is nothing but the 3D isotropic Ising lattice gauge Higgs model\cite%
{Fradkin_1979}, which can also be obtained from mapping the 2D quantum Hamiltonian 
(\ref{HTC_field}) onto a (2+1)-dimensional classical model based on imaginary time
evolution\cite{Gauge_Higgs_2010}.

In the absence of the field $h_z$, we have $K=0$ and the obtained partition
function is reduced to the 3D Ising gauge model\cite%
{lattice_gauge_theory_1979}:
\begin{equation}
\mathcal{Z}_{N}=\sum_{\sigma }\exp \left[ K_{\bm{p}}\sum_{\bm{p}%
}\prod_{\langle \bm{i}\bm{j}\rangle \in \bm{p}}\sigma _{\bm{i}\bm{j}}\right],
\end{equation}
and a 3D Ising type of phase transition occurs at a critical value of $K_{%
\bm{p}}$. On the other hand, along the $h_{z}$ axis, $K_{\bm{p}}=\frac{1}{2}%
\log \frac{1+h_{x}^{2}}{2h_{x}}$ is extremely large and the corresponding
partition function can be written
\begin{equation}
\mathcal{Z}_{N}=\sum_{\bm{\sigma}}\prod_{\bm{p}}\left( 1+\prod_{{\langle %
\bm{i}\bm{j}\rangle }\in \bm{p}}\sigma _{\bm{i}\bm{j}}\right) \exp \left[
K\sum_{\langle \bm{i}\bm{j}\rangle }\sigma _{\bm{i}\bm{j}}\right].
\label{temp}
\end{equation}
When new spin variables $\mu _{\bm{i}}$ are introduced on the cubic lattice sites 
$\bm{i}$, the spin variables $\sigma _{\bm{i}\bm{j}}$ on the bond between the
sites $\bm{i}$ and $\bm{j}$ can be represented by the new variables\cite%
{Ising_gauge_2007}: $\sigma _{\bm{i}\bm{j}}=\mu _{\bm{i}}\mu _{\bm{j}}$, and
the partition function of a 3D Ising model is thus obtained
\begin{equation}
\mathcal{Z}_{N}=\sum_{\mu }\exp \left[ K\sum_{\langle \bm{i}\bm{j}\rangle
}\mu _{\bm{i}}\mu _{\bm{i}}\right],
\end{equation}
which is a 3D Ising spin model on a cubic lattice. There is a ferromagnetic-paramagnetic 
phase transition at a critical value of $K$.

Moreover, when $h_{x}=h_{z}$, $K$ and $K_{\bm{p}}$ satisfies the self-dual
condition of the Ising gauge-Higgs model\cite{wegner_duality_1971}: $%
2K=-\log \tanh K_{\bm{p}}$, $2K_{\bm{p}}=-\log \tanh K$. In Fig. \ref%
{Gauge_higgs} (b), we displays the phase diagram of this 3D classical model
derived from the numerical Monte Carlo calculations\cite%
{Gauge_Higgs_1980,Gauge_Higgs_2010,youjin2012}, and the similar phase
diagram has been also obtained from higher-order perturbation expansion\cite%
{Vidal2009}. Actually the phase diagram is symmetric about the self-dual
line $h_{x}=h_{z}$, and the continuous phase transition between the toric
code phase and Higgs (confining) belongs to the 3D Ising universality class.
There is a first-order transition line along the self-dual line, but it ends
at a critical point. In addition, the three transition lines just meet at a
tri-critical point.

\textit{Conclusion and Outlook}. -We have generalized the framework of RK
wavefunctions using the TFD states. Compared with the RK wavefunctions, the
TFD states contain the information of both equal and unequal time
correlation functions. The partition function describing general quantum
phase transitions can be extracted from the R\'{e}nyi entropy of the TFD
state ansatz and it is further interpreted as a 3D classical model by using
the PEPS representation of the TFD state. These ideas are illustrated by
mapping the general deformed toric code wavefunction into the 3D classical $%
\mathbb{Z}_{2}$ gauge-Higgs model. Moreover, our framework might be useful
for studying the thermodynamical properties of quantum critical points. We
notice the PEPS method has been used to numerically study generic quantum
critical points\cite{Corboz_Finite_IPEPS_2018,Lauchli_Finite_IPEPS_2018},
which is beyond the RK type wavefunction with non-generic conformal quantum
critical points\cite{ardonne_2004}. The static critical exponents have been
accurately extracted with the finite correlation scaling method\cite%
{Schuch-2020}. However, the information of dynamical exponents is not
included in the PEPS. This might be improved by taking the TFD states into
consideration.

\textit{Acknowledgements.} -The authors are grateful to Qi Zhang for the
stimulating discussions. The research is supported by the National Key
Research and Development Program of MOST of China (2017YFA0302902).

\bibliography{3d_model_ref}

\end{document}